# A New Classification of Clustering-based for Different Problems in Different Wireless Ad-hoc Networks

Adda Boualem[1] and Marwane Ayaida[2] and Hichem Sedjelmaci[3] and Chaimaa Khalfi[1] and Kamilia Brahimi[1] and Bochra Khelil[1] and Sanaa Bouchama[1]

[1]Department of Computer Science, Ibn Khaldoun
adda.boualem@univ-tiaret.dz, chaimaa.khalfi@univ-tiaret.dz, kamilia.brahimi@univ-tiaret.dz, bochra.khelil@univ-tiaret.dz, Sanaa.Bouchama@univ-tiaret.dz

[2] IEMN UMR CNRS 8520, Université Polytechnique Hauts-de-France 59300 Valenciennes, France
marwane.ayaida@univ-uphf.fr

[3]Ericsson R&D Security Massy Palaiseau, France
hichem.sedjelmaci@ericsson.com

## Abstract

*Ad-hoc networks are specifically designed to facilitate communication in environments where establishing a dedicated network infrastructure is exceedingly complex or impractical. The integration of clustering concepts into various ad-hoc network scenarios, including Wireless Sensor Networks (WSN), Mobile Ad-hoc Networks (MANET), Vehicular Ad-hoc Networks (VANET), Delay-Tolerant Networks (DTN), Wireless Ad-hoc Networks (WANET), Underwater Wireless Sensor Networks (UWSN), Unmanned Aerial Vehicle Networks (UAV), commonly known as "drones," Space Networks (SN), and Satellite Networks (TN), presents abundant opportunities for refining strategies in event tracking and area monitoring across both deterministic and uncertain environments. This paper conducts a comparative analysis of diverse proposed strategies leveraging clustering concepts to address coverage challenges within deterministic and uncertain environments. As a result, it addresses current and future challenges inherent in clustering-based WANET, elucidating the merits, shortcomings, and weaknesses of clustering models. Lastly, it identifies potential avenues for addressing coverage issues in existing and emerging technologies.*

## 1. Introduction

The ad-hoc networks have been particularly designed for establishing communications in an ad-hoc network is created using a group of wireless nodes that can communicate with one another. Wireless Ad-hoc network application research is multidisciplinary, including data analysis, economics, mathematics, forensics, information technology, and computer science. This context covers cutting-edge research in wireless network-based clustering from diverse environments and fields on the complex subject of wireless communications uses. Actually, technological advancements in Ad-hoc wireless networks and mobile devices are revolutionizing the reliability of capturing, capturing, and communicating environmental information. The real model is limited by various uncertainties, such as the inaccuracy of measurement components, atmospheric phenomena, intrusions, natural phenomena such as animals, volcanoes, rivers, . . . industrial phenomena, etc. [1]. The research by Boualem [1] focuses on improving coverage in wireless sensor networks, a topic that is crucial for various applications, including healthcare and agriculture. Today, inexpensive smart sensors networked through wireless connections and deployed in large numbers offer unprecedented opportunities to monitor homes, cities, and environments. Furthermore, networked sensors offer a broad spectrum of defence applications, ushering in new capabilities for reconnaissance, surveillance, and various tactical uses. A range of studies has explored the use of fuzzy logic in improving coverage and query processing in wireless sensor networks. [2] Proposed a fuzzy/possibility approach for area coverage, while [3] and [5] both introduced fuzzy-based approaches for improving network coverage and target coverage, respectively.

Focusing on the ability of devices to detect physical phenomena such as moisture, heat, and pressure is a very important practical task for wireless sensor networks. Specify the deterministic and non-deterministic effects that affect these capabilities in deterministic and non- deterministic environments (hostility of the environment in which sensor nodes) are deployed due to changes in atmospheric conditions and modifications to the deployed sensor network topology. Unreliability of communication stations and receiving stations etc. affects the quality of service and decisions related to real information [10].

Wireless ad-hoc networks are networks that are formed spontaneously and without the need for any pre-existing infrastructure or centralized management. Ad-hoc networks can face several issues; Connectivity Issues (Obstacles or interference can affect connectivity), Dynamic Topology (Ad-hoc networks have a dynamic topology, with devices joining and leaving the network frequently), Security Concerns (Ad-hoc networks are more vulnerable to security threats due to their decentralized nature), Limited Bandwidth (Ad-hoc networks may have limited bandwidth, especially in scenarios with many connected devices), Power Consumption, Scalability, Interference, and the Quality of Service (QoS), . . .

The main objective is to identify the optimal classification of efficient clustering methods for addressing the myriad challenges inherent in ad-hoc networks in real-world settings.

As we are concerned with clustering solutions in almost ad-hoc network issues and challenges, we will propose an efficient classification-based clustering for almost ad-hoc network problems, and we will review existing clustering in ad-hoc network issues.

The remainder of this paper is organized as follows: Section 2 presents some fundamental concepts of wireless Ad-hoc networks. Section 3 reviews the problems that impact the Wireless Sensor Networks and shows the relationship between these problems. In Section 4, an exploration of related works is undertaken, with a bifurcation into two principal segments. The initial portion scrutinizes classical clustering-based Wireless Ad-hoc Networks, outlining models and elucidating limitations to catalyze novel research trajectories. Conversely, the subsequent segment delineates our classification clustering-based wireless network models, pinpointing shortcomings to spur fresh avenues of inquiry. A comprehensive comparative analysis spanning wireless network types, clustering methodologies, node classifications, attained objectives, and the utilization of certain/uncertain methods or environments is provided. Proceeding to Section 5, an exposition on the present and forthcoming challenges pertaining to Coverage is provided. Finally, Section 6 furnishes a conclusion, encapsulating insights and charting paths for future endeavors.

## 2. Fundamental Concepts of Wireless Ad-Hoc Network

[4], [5] Provides a comprehensive overview of wireless ad-hoc networks, including their types, characteristics, differences, applications, and protocols. Among these types, we compared these types in terms of type, characteristics, type of nodes, and the purpose of each type of WANET illustrated in Table 1.

## 3. Networks-based Coverage Problems

One of the main challenges of wireless networks is coverage, which consists of continuously and efficiently observing phenomena or events that may occur in a geographic area. The coverage problem has been studied for over a decade as a mathematical problem.

Table 1: Wireless Ad-hoc Networks comparison [6]

| Wireless Network Type | Nodes Type | Characteristics | Depl. Envir | Objectives |
|---|---|---|---|---|
| WSN | Static or, and Mobile sensor nodes | serves to relay sensor data from the physical environment to a central location | 2D, 3D | Gather information about physical, environmental, or relevant conditions, distributing the data spatially. Following data collection, the amassed data is transmitted to a central system, facilitating the measurement of vital environmental variables including temperature, sound levels, pollution levels, and other pertinent factors. |
| WANET or MANET | Mobile Nodes | A decentralized type of wireless network. The network is ad hoc because it does not rely on a pre-existing infrastructure like routers or wireless access points | 2D, 3D | Determining which nodes forward data is made dynamically based on network connectivity and the routing algorithm, each node participates in routing by forwarding data to other nodes |
| VANET | Vehicles | Utilizes the radio waves, vehicles establish instantaneous communication networks on the fly as they navigate along roads | 2D | Serve as a means of communication between vehicles and roadside equipment. Intelligent Vehicular Ad hoc Networks (In VANETs) represent a form of artificial intelligence that enhances the ability of vehicles to exhibit intelligent behavior in situations such as vehicle-to-vehicle collisions and accidents, and it is imperative to ensure the security of VANETs through the implementation of lightweight protocols |
| DTN | Static and Mobile nodes | Unlike traditional networks that assume continuous end-to-end connectivity, Delay-Tolerant Networks (DTN) is a type of network designed to operate effectively in situations where network connectivity is intermittent or unreliable. Key characteristics and | 2D or 3D | Designed to accommodate situations where continuous connectivity between nodes is not guaranteed, allowing for delays in message delivery), Routing Protocols (DTNs often employ specialized routing protocols that consider the dynamic and intermittent nature of the network, optimizing message delivery under challenging conditions) |

| | | features of Delay-Tolerant Networks include; Intermittent Connectivity | | |
|---|---|---|---|---|
| UWSN | Dynamic and Static nodes | Are specialized networks designed for communication and data exchange in underwater environments | 3D | These networks typically consist of a collection of underwater sensor nodes that work collaboratively to collect and transmit data from the aquatic environment |
| UASN | Dynamic and Static nodes | UASNs are specialized networks designed for communication and data exchange in underwater environments using acoustic signals | 3D | These networks leverage acoustic waves to transmit information between underwater sensor nodes, enabling a variety of applications in oceanography, and environmental monitoring |
| UAV | Mobile and Static | Refer to systems of unmanned aerial vehicles (UAVs) that are interconnected to enable communication, coordination, and collaborative operations | 2D | These networks play a crucial role in various applications, including surveillance, reconnaissance, agriculture, disaster response, and scientific research |
| TN | Dynamic and Static nodes | Refers to communication systems designed for military operations, providing reliable and secure connectivity in dynamic and challenging environments | 2D and 3D | These networks play a critical role in facilitating communication among military units, enhancing situational awareness, and supporting the coordination of tactical activities |

## 4. Clustering-based Wireless Ad-hoc Networks

In this section, we will elucidate our primary contributions: (a) introducing a novel classification for clustering- based Wireless Ad-hoc Networks, and (b) presenting a generalized pseudo-algorithm for cluster construction.

Various studies have delved into the exploration of wireless ad-hoc clustering. [17] introduced an innovative cluster-based architecture designed for cognitive wireless ad hoc networks, with a specific emphasis on improving data forwarding efficiency and load balance. [18] proposed the Mobility-Aware Pro-Active Low Energy (MAPLE) clustering scheme, surpassing existing access-based clustering protocols in terms of stability, overhead, and load distribution. [19] presented a mobility-based clustering approach that organizes mobile nodes into adaptive variable- sized clusters based on their mobility, utilizing both physical and logical network partitions. [20] devised a contention-based clustering algorithm tailored for wireless ad hoc networks, showcasing advantages in scalability and stability, particularly in large networks.

In the following section, we will describe; (a) the classical clustering-based Wireless Ad-hoc Networks: classification, and (b) our new Clustering-based Wireless Ad-hoc Networks: classification.

### 4.1. Classical Clustering-based Wireless Ad-hoc Networks: classification

In literature, there are no more classifications. Among them, we quote the classification based on frequency [21] (Figure 3). [1][7][8][9], [9] and [23] propose a new classification of coverage types in WSN (Figure 1, Figure 2, Figure 4). However, there is a rich literature on works based on clustering as a solution for various problems to be addressed in Wireless Ad-hoc Networks: (Coverage, energy, routing, connectivity, security, quality of service …

#### 4.1.1. Our classification of Uncertain-based coverage models

Wireless Ad-hoc Networks: are distinguished by their adaptability and agility, crucial for wireless communications, and possess the capability to maintain connectivity for individuals even as they move between locations. Our objective in this paper consists of two contributions; (a) we have proposed a new classification of the necessity of clustering to trial problems and challenges for all Wireless Ad-hoc Networks, and (b) a pseudo-algorithm; is a basic pseudo-algorithm for constructing clusters in a wireless ad- hoc network (Part 1: Algorithm 1), and (Part 2: Algorithm 2). This classification as shown in Figure 5 is divided into two main axes: (a) a deterministic-based clustering classification and the other is uncertainty-based clustering, where the employed strategies are deterministic or uncertain in deterministic or uncertain environments, respectively. Table 2 displays a compilation of related works focusing on clustering-based approaches tailored for addressing various problems across diverse Wireless Ad-hoc Networks, where D and U mean deterministic-based clustering and uncertain-based clustering respectively.

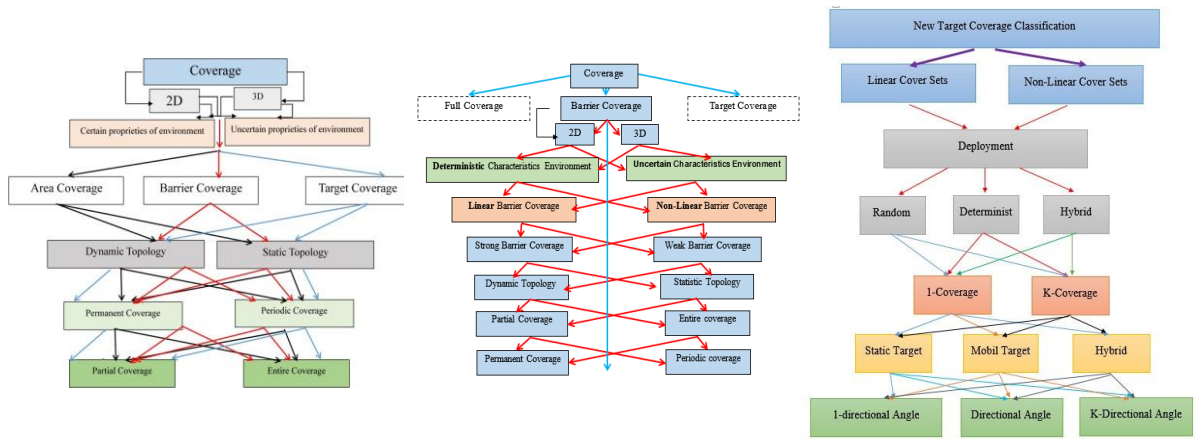

Figure 1: Classification of Coverage, Barrier and Target Coverage Types in WSN [1], [7], [8], [9]

## 5. Clustering- Current and future challenges of Clustering-based Wireless Ad-hoc Networks

In Wireless Ad-hoc Networks: clustering refers to the grouping of nodes into clusters to improve network efficiency, management, and communication. However, the process of clustering comes with its own set of challenges and problems. Some challenges and issues of WANETs- based clustering:

- Cluster Formation: WANETs often have a dynamic and unpredictable topology due to

node mobility. Forming stable clusters in such an environment is challenging.
- Cluster Head Selection: The selection of an energy- efficient cluster head is crucial for prolonging the network's lifetime. Achieving a balance in energy consumption among nodes poses challenges.
- Cluster Maintenance: Among the challenges of WANET are;
  - The failure or departure of a cluster head may disrupt the cluster structure, requiring efficient mechanisms for re-election or replacement.
  - Addressing node mobility and ensuring that the cluster structure adapts to changes in node positions is a non-trivial task.
  - Efficient communication within clusters is essential. Overhead in managing intra-cluster communication and avoiding congestion is a challenge.
  - Optimizing communication between clusters, especially in the presence of multiple clusters, is challenging due to the lack of a fixed infrastructure.

Security and Privacy: Ensuring the security of cluster heads is crucial as they play a central role in managing cluster activities. Protecting against attacks or compromised cluster heads is a challenge, and Maintaining data privacy within clusters, especially when nodes share sensitive information, requires robust encryption and authentication mechanisms.

- Resource Constraints: WANETs often have limited bandwidth. Efficiently utilizing available bandwidth for both control and data messages in a clustered environment is challenging, and the clustering should aim at energy-efficient communication and data processing to extend the network's lifetime, but achieving this balance is challenging.
- Scalability and Quality of Service: Extending clustering mechanisms to large-scale networks introduces scalability challenges. Efficient cluster management becomes more complex as the network size increases, and providing reliable and timely communication services, especially in applications with stringent QoS requirements, poses challenges in clustered WANETs. Moreover, recovering from cluster head failures or node departures without significant disruption to the network's operation is a challenge.
- Different applications may have diverse requirements, making it challenging to design clustering mechanisms that cater to the varied needs of different use cases.

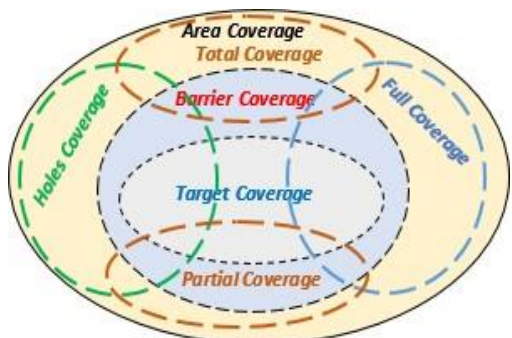

Figure 2. Classification of Coverage Types in WSN. [9]

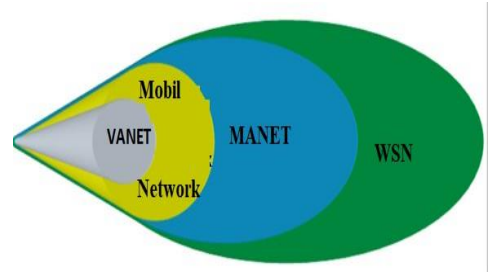

Figure 3. Classical Classification of Ad-Hoc Mobil Networks. [23]

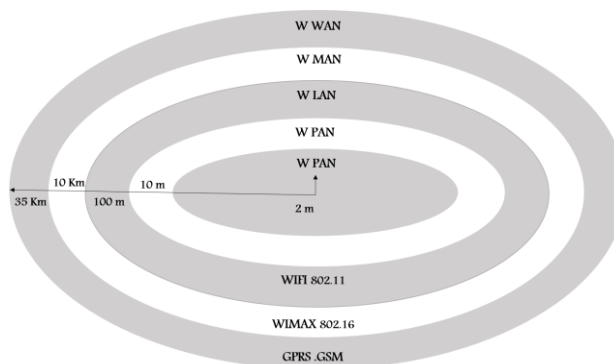

Figure 4. Classification of Wireless Ad-hoc Networks: with their signal range. [21]

Table 2: Some work on Wireless Ad-hoc networks based on clustering

| Ref | Netw Type | Highlight | Used Technique | Objectives |
|---|---|---|---|---|
| [10] | WSN | Clustering the network enhances energy efficiency by enabling data sharing among cluster members, negating the necessity for data transmission across multiple nodes. | Employing an AI-based approach, clustering-based techniques integrated with Self-Organizing Map (SOM) are utilized to conserve energy in resource-constrained networks. | Excels in terms of minimizing energy consumption and addressing computational challenges within the network |
| [11] | VANET | Suggest rapid randomized algorithms designed for clustering in VANETs, with a specific emphasis on prioritizing 2-hops clustering. | Implement HCA and HCA-2 algorithms for facilitating 2-hops clustering in vehicular ad-hoc networks | Both HCA and HCA-2 algorithms are specifically crafted to manage channel access and scheduling within vehicular networks, independent of vehicle localization |
| [12] | VANET | These investigations have established a classification system for clustering in VANETs, delineating numerous challenges and corresponding solutions. | Employing clustering within vehicular ad hoc networks (VANETs) showcases its capability to mitigate challenges like scalability and stability. | This study has concentrated on categorizing clustering algorithms into intelligence-based, mobility-based, and multi-hop-based classifications. |
| [13] | MANET | Offers a thorough examination of clustering schemes in MANETs, organizing them according to criteria for Cluster Head (CH) selection and evaluating their operational performance. | Use a self-organization-based clustering scheme using zone-based group mobility. | Improves scalability and stability. |
| [14] | WSN | Categorizing strategies into classical, optimization, and machine learning techniques | Delves into the enhancement of routing-based clustering methodologies, with a particular emphasis on multi-hop routing. It categorizes techniques into parameter-based, optimization-based, and | An exhaustive examination of clustering within wireless sensor networks is provided. |

| | | | methodology-based approaches for comprehensive exploration. | |
| --- | --- | --- | --- | --- |
| [15] | MANET | The mimetic algorithm-based clustering technique demonstrates superior performance compared to existing methods concerning cluster counting, reaffiliation rate, cluster longevity, and control message overload. | Introduces memeHoc, a algorithm-based clustering for MANETs Memetic technique | This work aims to reduce topology maintenance message overload |
| [16] | UWSN | An SS-GSO approach is proposed to enhance energy efficiency and equilibrium in clustering algorithms for underwater wireless sensor networks | The proposed SS-GSO method improves upon existing techniques by constructing a comprehensive fitness function that incorporates diverse sources of information, such as total energy, residual energy, and Lucifer in value. | Prolong the network's longevity while maintaining equilibrium in energy consumption and reducing transmission distances. |

## 3. CONCLUSIONS

In conclusion, this research paper has introduced a novel classification framework for wireless Ad-hoc networks based on clustering, elucidating its advantages and identifying potential future challenges. This classification not only enhances our understanding of the diverse landscape of wireless networks but also provides a structured approach to categorizing them based on clustering methodologies.

The advantages of this classification lie in its ability to offer a clear taxonomy for various types of wireless networks, facilitating improved comprehension and comparison among different clustering-based approaches. It underscores the significance of clustering in wireless network design, management, and performance optimization. The insights gained from the research contribute to the advancement of wireless network architectures by addressing current limitations and laying a foundation for further innovations.

Looking ahead, several challenges emerge. The dynamic nature of wireless environments, the scalability of clustering techniques, and the need for robust security measures present ongoing areas of concern. Future research should focus on developing more adaptive and scalable clustering algorithms to meet the evolving demands of wireless networks.

Moreover, the integration of emerging technologies, such as machine learning and artificial intelligence, holds promise for enhancing the efficiency and adaptability of clustering-based wireless networks. This paper sets the stage for future endeavors, encouraging researchers and practitioners to explore innovative solutions to overcome challenges and unlock the full potential of clustering in the realm of wireless networks.

In summary, while this research provides valuable insights and advancements in the classification of wireless networks based on clustering, it also highlights the need for ongoing research and development to address emerging challenges and opportunities.

---

**Algorithm 1** Basic Pseudo-Algorithm for design a Cluster set in an Ad-hoc Wireless Network: Part1

1: **Initialization step**
**Require:** Set a timer (t=N);   //for cluster formation
**Require:** Define parameters;   //such as transmission range, cluster head selection criteria,...
**Require:** Each node is initially in its own cluster.
2: **Step 1: Discovery Phase**
3: Nodes broadcast HELLO messages to discover neighboring nodes within their transmission range.
4: Nodes collect information about their neighbors, including their IDs and signal strengths.
5: **Neighbor Table Construction**
6: Each node constructs a neighbor table based on the received HELLO messages.
7: The table includes information about neighboring nodes, such as ID, signal strength, and other relevant metrics.
8: **Step 2: Cluster Head Election**
9: Nodes evaluate their candidacy $Cnd(s_i)$ for being a cluster head $(CH)$ based on predefined criteria //(e.g., highest residual energy, better connectivity)
10: **if** $(Cnd(s_i) > Cnd(s_j), j \neq i)$ **then**   //Nodes with higher candidacy become cluster heads
11:     $C_k^i = \{C_k^i \bigcup s_j\}$;   //Non-cluster head nodes join the cluster of the nearest cluster head
12: **end if**
13: **Step 3: Cluster Formation**
14: Send(CH-message, Id-CH);   //Cluster heads broadcast a message to announce their presence and cluster Id
15: **while** $((s_j \neq CH)$ and $(d(s_j, CH) \leq R_s(n_i))$ **do**
16:     $C_k^i = \{C_k^i \bigcup s_j\}$;   //Non-cluster head $s_j$ nodes join the cluster $C_k^i$ with the nearest cluster head
17:     **if** $CH \lll 0$ or $E(CH) \ll$ **then**   //a cluster head becomes inefficient $(CH \lll 0)$ or its energy level is low $(E(CH) \ll)$
18:         A new election may be triggered;
19:     **end if**
20: **end while**
21: **Step 4: Data Transmission**
22: Nodes within a cluster communicate with their cluster head for data transmission.
23: Cluster heads manage inter-cluster communication and forward data between clusters.
24: **Step 5: Periodic Update step**
25: **for** t=N to 0 **do**   //for a period of time t
26:     Nodes reassess their candidacy for cluster-heads based on the current state; State Adjusting for factors like energy consumption;
27:     **if** $CH \lll 0$ or $E(CH) \ll$ **then**   //a cluster head becomes inefficient $(CH \lll 0)$ or its energy level is low $(E(CH) \ll)$
28:         A new election may be triggered;
29:     **end if**
30: **end for**

---

**Algorithm 1** Basic Pseudo-Algorithm for design a Cluster set in an Ad-hoc Wireless Network: Part2

1: **Step 6: Dynamic Adjustment**
2: **for** t=N to 0 **do**   //for each period t =N
3:     **for** C=1 to n **do**   //for cluster head based on the current state
4:         Nodes reassess their candidacy;
5:         Adjusting for factors;   //like energy consumption;
6:         Nodes monitor the network conditions;
7:     **end for**
8: **end for**
9: **if** Significant changes occur in the network **then**   //e.g., node failure, new nodes entering the network
10:     Reconfigure the clusters;
11:     Reselect cluster heads or redistributing nodes among clusters;
12: **end if**
13: **Step 7: Stop**
14: **if** The clustering process continues **then**
15:     set $t = N$;
16:     goto step 1;
17: **else**
18:     Terminate the algorithm;
19: **end if**

---

**Algorithm 2** Typical Pseudo-Algorithm for building Cluster set in an Ad-hoc Wireless Network

**Require:** $n$ the number of wirless Ad-hoc nodes to be deployed;
**Require:** $k$ the number of CHs to be used in the Wireless Ad-hoc Network;
**Require:** $C_k^i$: Node $i$ matches the set of nodes $C_k$ neighboring of the $CH_k$ ;
**Require:** $d(x, y)$ is the distance between $x$ and $y$;
**Require:** $X$ is a boulean variable that indicates whether node $i$ can be included in the neighboring set of $CH$;
**Require:** $X = 0$;
1: **for** j=1 to k **do**
2:     **for** i=1 to n **do**
3:         **if** $d(s_j, CH_i) < R_s$ **then**   //$s_j$ at monitoring range of Cluster-Head $CH_i$
4:             $C_k^i = \{C_k^i \bigcup s_j\}$;   //$C_k^i$ is the neighboring set of $CH_k$ to be covered by the sensor $i$
5:             $X = 1$;
6:         **end if**
7:         **if** $X = 0$ **then**   //No sensor nodes in the vicinity
8:             Deploy a sensor node $t_k$ where $d(t_k, s_m) < R_s$   //$t_k$ at monitoring range of sensor node $m$
9:         **end if**
10:         $X = 0$;
11:     **end for**
12: **end for**

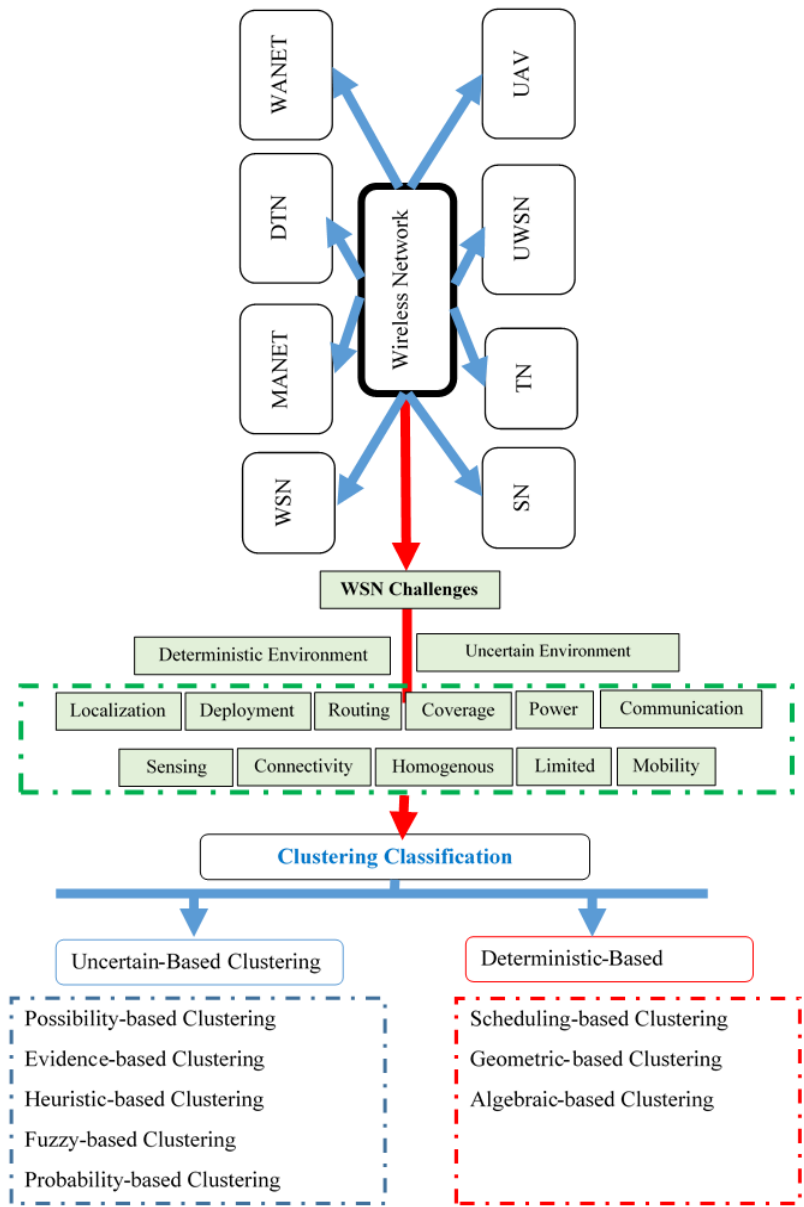

Figure 5. The Pipeline of the New Classification of Clustering-based for Different Problems in Different Wireless Networks

## ACKNOWLEDGEMENTS

The authors would like to thank everyone, just everyone!

## AUTHORS

**Adda Boualem** is an Associate Professor, **Chaimaa Khalfi, Kamilia Brahimi, Bochra Khelil and Sanaa Bouchama** are Master's students in Department of Computer Science at University of Tiaret, Algeria.

**Marwane Ayaida** is professor at Université Polytechnique Hauts-de-France 59300 Valenciennes, France.

**Hichem Sedjelmaci** is a proEricsson R&D Security Massy Palaiseau, France

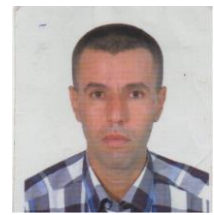